\documentstyle[prb,aps,multicol,psfig]{revtex}

\newcommand{\mrm}[1]{{\rm{#1}}}
\newcommand{\Pcav}{P_\mrm{cav}}
\newcommand{\Ps}{P_\mrm{s}}
\newcommand{\Pst}{P_\mrm{stat}}
\newcommand{\Pmax}{P_\mrm{max}}
\newcommand{\Pmin}{P_\mrm{min}}
\newcommand{\etal}{\textit{et~al.}}
\newcommand{\Vc}{V_\mrm{c}}

\begin{document}
\draft
\title{Cavitation pressure in liquid helium}
\author{F.~Caupin and S.~Balibar}
\address{Laboratoire de Physique Statistique de l'Ecole Normale Sup\'erieure associ\'e aux Universit\'es Paris 6 et Paris 7 et au CNRS,\\
24 Rue Lhomond, 75231 Paris Cedex 05, France}
\date{19 July 2001}
\maketitle

\begin{abstract}
{\small Recent experiments have suggested that, at low  enough temperature, the homogeneous nucleation of bubbles occurs in liquid helium near the calculated spinodal limit. This was done in pure superfluid helium~4 and in pure normal liquid helium~3. However, in such experiments, where the negative pressure is produced by focusing an acoustic wave in the bulk liquid, the local  amplitude of the instantaneous pressure or density is not directly measurable. In this article, we present a series of measurements as a function of the static  pressure in the experimental cell. They allowed us to obtain an upper bound for the cavitation pressure $\Pcav$ (at low temperature, $\Pcav < -2.4 \,\mrm{bar}$ in helium~3, $\Pcav < -8.0 \,\mrm{bar}$ in helium~4). From a more precise study of the acoustic transducer characteristics, we also obtained a lower bound (at low temperature, $\Pcav > -3.0 \,\mrm{bar}$ in helium~3, $\Pcav > - 10.4 \,\mrm{bar}$ in helium~4). In this article we thus present quantitative evidence that cavitation occurs at low temperature near the calculated spinodal limit ($-3.1 \,\mrm{bar}$ in helium~3 and $-9.5 \,\mrm{bar}$ in helium~4). Further information is also obtained on the comparison between the two helium isotopes. We finally discuss the magnitude of nonlinear effects in the focusing of a sound wave in liquid helium, where the pressure dependence of the compressibility is large.}
\end{abstract}
\pacs{PACS Numbers: 67.55.Cx, 67.40.Kh, 64.60.Qb, 43.25.+y}

\begin{multicols}{2}
\section{Introduction}

As has often been explained, the purity of liquid helium offers unique opportunities to study phase transitions.\cite{Balibar98,Balibar00} Furthermore, liquid helium is a simple system and its thermodynamic properties have been accurately measured, so that several authors\cite{Solis92,Guilleumas93,Boronat94,Maris94,Maris95,Dalfovo95,Campbell96,Bauer00,Casulleras00} have calculated the extrapolation at negative pressure of its equation of state. All these calculations are reasonably consistent with each other; they predict the existence of a spinodal limit $\Ps$ at $- 3.1 \,\mrm{bar}$ for liquid helium~3 and at $-9.5 \,\mrm{bar}$ for liquid helium~4, near the absolute zero.\cite{Mariserror} As one approaches the spinodal limit, the compressibility diverges so that the energy barrier for the nucleation of bubbles in the stressed liquid vanishes, and the liquid becomes totally unstable.

In order to test this prediction, we have studied cavitation by focusing a high intensity acoustic wave in bulk liquid helium, far from any wall. If one wants to approach the spinodal limit, it is necessary to work at very low  temperature, otherwise thermal fluctuations allow the system to pass a rather high energy barrier at higher, i.e., less negative, pressure. This is possible in liquid helium only, since no other liquid exists down to zero temperature. Eventually, close enough to the spinodal limit and at low enough temperature, a crossover has been predicted to exist from a thermally activated classical cavitation to a quantum regime where the nucleation of bubbles occurs by the quantum tunneling of a sizable quantity of liquid.\cite{Maris95,Guilleumas96}

In previous experiments, we obtained some evidence for the existence of this crossover in helium~4. Below about $0.6 \,\mrm{K}$, Lambar\'e~\etal\cite{Lambare98} observed a stochastic and temperature independent behavior, while above $0.6 \,\mrm{K}$ they observed another stochastic behavior where the cavitation threshold decreased with temperature. These observations agreed with the theoretical predictions by Maris\cite{Maris94,Maris95} if the quantum cavitation occurred at $-9.2 \,\mrm{bar}$, close enough to his calculated value of the spinodal limit. This agreement was obtained after noticing that, given the amplitude of the wave, and due to adiabatic cooling during the negative pressure swing, the minimum instantaneous temperature was lower than the static temperature by a factor of $3$ (that is, $0.2 \,\mrm{K}$ for a static temperature of $0.6 \,\mrm{K}$). Unfortunately, in this experiment, it was not possible to measure directly the exact value of the negative swing in the acoustic wave when nucleation occurs. Instead we measured the voltage applied to the transducer that generates the wave; as for the wave amplitude in the focal region, it is difficult to estimate since one expects the focusing of the wave to be nonlinear and its precise calculation in a cylindrical geometry has not yet been done.

Given these results and difficulties, we tried a comparison with helium~3. We first confirmed that the cavitation threshold pressure is less negative in helium~3 than in helium~4.\cite{Caupin98} Moreover, we did not observe a crossover to a temperature independent quantum regime of cavitation. The latter observation may be understood by invoking theoretical arguments involving dissipation\cite{Jezek99,Burmistrov00} or Fermi liquid properties.\cite{Balibar98,Caupin01Fermi}

In this context, we have tried to improve our estimation of the cavitation pressure by carefully measuring the dependence of the cavitation voltage on the static pressure $\Pst$ in the cell. As explained in this article, this study allows us to present an upper bound for the cavitation pressure in both helium~3 and helium~4 ($\Pcav < -2.4 \,\mrm{bar}$ in helium~3, $ \Pcav < - 8.0 \,\mrm{bar}$ in helium~4). From a more
\begin{figure}
\begin{center}
\psfig{file=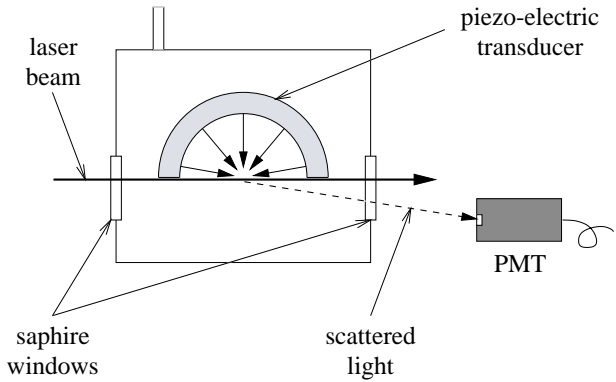,height=5cm}
\end{center}
\caption{Sketch of the experimental setup: a high amplitude pressure swing is generated at the focus of the hemispherical transducer; a laser beam passes through the acoustic focal region, and the light scattered by cavitation is detected with a photomultiplier tube (PMT).}
\label{fig:setup}
\end{figure}
{\noindent}precise study of the acoustic transducer characteristics, we also obtained a lower bound ($\Pcav > -3.0 \,\mrm{bar}$ in
helium~3, $ \Pcav > - 10.4 \,\mrm{bar}$ in helium~4). Finally, we also encountered interesting questions about the magnitude of nonlinear effects in the focusing of high intensity acoustic waves in a medium such as liquid helium where the compressibility strongly depends on pressure.

\section{Experimental procedure}

\subsection{Experimental setup}

As shown in Fig.~\ref{fig:setup}, our experimental setup is similar to the one used in previous experiments.\cite{Lambare98,Caupin98,Caupin00} We have reduced the inner volume of the cell to $4.5 \,\mrm{cm^3}$ in order to lower the cost of the helium~3 experiment. This cell is anchored to the mixing chamber of a dilution refrigerator by copper columns with gold plated contacts. Carbon resistor thermometers and a heater are thermally connected to the outside of the copper cell walls. A burst of $1\rm\,MHz$ ultrasound is emitted and focused in the liquid by a hemispherical piezoelectric transducer (the ``ceramic''). The cell is connected to a buffer volume at room temperature, so that we easily monitored the static pressure in the cell. When working with helium~3, this was done with a capacitive sensor (Keller type PAA-41) with an accuracy of $\pm 0.5 \,\mrm{mbar}$. For helium~4, we used conventional pressure gauges; the uncertainty was $\pm 0.7 \,\mrm{mbar}$ for low pressure measurements and $\pm 12.5 \,\mrm{mbar}$ for static pressures above $1 \,\mrm{bar}$. Two sets of four windows allow us to shine a laser beam from the outside of the cryostat through the acoustic focal region, and to detect the light that is scattered by local changes in density. The detector is a photomultiplier tube with a response time shorter than $0.5 \,\mrm{\mu s}$. 
The acoustic wave scatters the laser light at small angle; when cavitation occurs, gas bubbles scatter light at a larger angle. By adjusting the photomultiplier
\begin{figure}
\begin{center}
\psfig{file=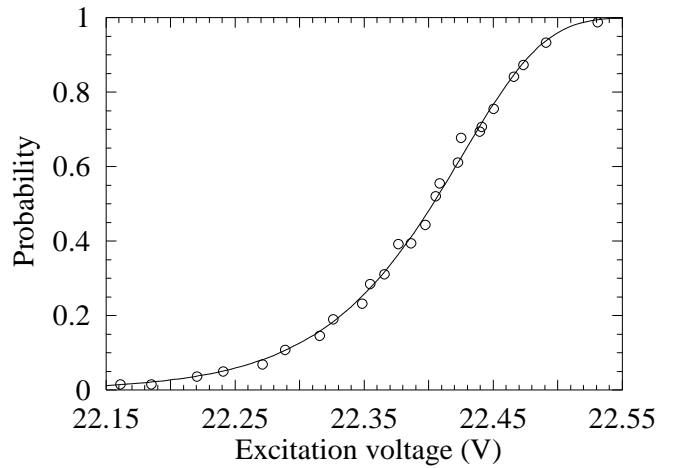,width=8.6cm}
\end{center}
\caption{Cavitation probability versus driving voltage in helium~3 at $T=566 \,\mrm{mK}$ and $P=4 \,\mrm{mbar}$. Each circle is a probability measurement over $1600$ bursts at a given voltage. The solid line is a fit with Eq.~(\ref{eq:S(V)}) which gives $\Vc=22.403 \,\mrm{V}$ and $\xi=355$.}
\label{fig:Scurve}
\end{figure}
{\noindent}tube position and the size of a diaphragm in front of it, we can choose to detect cavitation either from the scattered light or from the light missing in the transmitted beam.

\subsection{Voltage measurements}

One side of the ceramic is grounded in the cell and the other is connected to  the output of a rf amplifier at room temperature. In order to improve the accuracy of the  probability measurements, we needed to increase the number of events to count. Furthermore, to study helium~3 down to $40 \,\mrm{mK}$, we had to reduce the dissipation by lowering the repetition rate of the acoustic pulses. Indeed, inside the ceramic itself, there is a mechanical dissipation of order $2.6 \,\mrm{\mu W}$ for six-cycle pulses repeated at $1 \,\mrm{Hz}$, a non-negligible amount. As a consequence, we had to improve the stability of the excitation. Since this piezoelectric transducer has a low impedance at resonance ($12$ to $16 \,\mrm{\Omega}$, see below), we had to minimize the possible drift of the impedance of its connecting cable. The previous cable included a $1.5 \,\mrm{m}$ section from $4 \,\mrm{K}$ to room temperature, that was a commercial cable made of stainless steel. We replaced this section by a homemade coaxial cable consisting of a $0.24\rm\,mm$ diameter copper wire separated from stainless steel tubing ($1.2 \,\mrm{mm}$~i.d., $1.5 \,\mrm{mm}$~o.d.) by a Teflon tube ($0.30 \,\mrm{mm}$~i.d., $0.76 \,\mrm{mm}$~o.d.). We kept the superconducting coaxial cable connecting the transducer to the $4 \,\mrm{K}$ region. The impedance of the whole line is now negligible compared to the transducer impedance, so that the excitation does not depend on the level of helium in the $4 \,\mrm{K}$ bath. We finally improved the stability of the excitation voltage itself by using a new generator (Hewlett Packard model 33120A). Our homemade rf amplifier uses an Apex PA09 circuit and is located in a thermally regulated box as before.\cite{Lambare98}

During the experiment, the ceramic is driven by a burst with a few cycles of a $1 \,\mrm{MHz}$ sine wave. This burst is monitored with a digital oscilloscope (Tektronik model TDS 420A, $200 \,\mrm{MHz}$, $100 \,\mrm{MS\, s^{-1}}$). The waveform is transferred to a computer using a  conventional IEEE interface and then fitted to a sine wave using the Levenberg-Marquardt method\cite{NumRec} with four adjustable parameters (amplitude, frequency, phase, and offset). The first cycle is always slightly distorted and therefore ignored in the fit. We have checked the distribution of amplitudes over $10^5$ fits under the same experimental conditions. The repetition rate being $1 \,\mrm{Hz}$ in this case, the total acquisition time was about $28\,\mrm{h}$. Over such a duration, we achieved a stability of $1.8 \times 10^{-3}$.

\subsection{Statistics of cavitation}

At a given temperature and pressure in the cell, we measured the cavitation probability $\Sigma$ over series of bursts at several driving voltages. As previously reported\cite{Lambare98,Caupin98} and illustrated in Fig.~\ref{fig:Scurve}, cavitation is a stochastic process, and $\Sigma(V)$ is well described by the ``asymmetric S-curve formula''
\begin{equation}
\Sigma(V) = 1 - \exp\left[- \xi \ln2 \,\exp\left(\frac{V}{\Vc} - 1 \right)\right] \; .
\label{eq:S(V)}
\end{equation}

We previously explained\cite{Lambare98} that this behavior is expected if cavitation is thermally activated. It is derived from a linear expansion of the activation energy $E$ around the cavitation threshold voltage $\Vc$ where the probability is $1/2$. The quantity $\xi$ is the inverse width of the distribution of events, given by $\xi=(\Vc/k_B T)(\partial E/ \partial V)$. A similar formula is also expected in the quantum regime: $\xi=(\Vc/\hbar)(\partial B/ \partial V)$ where $B$ is an action. In our whole study, the cavitation threshold $\Vc$ is determined from a fit of $\Sigma$ with Eq.~(\ref{eq:S(V)}); the accuracy is $\pm 0.1\,\%$.

\subsection{Piezoelectric transducer characteristics}

\label{par:ceram}
In the following discussion, we will need the various parameters that describe the transducer behavior. It is made of lead zirconium titanate (Quartz \& Silice type P7-62); its thickness is $2 \,\mrm{mm}$, its inner radius $R_{\rm{trans}}$ is $8 \,\mrm{mm}$, and its mass $M$ is $7.5 \,\mrm{g}$. It is used at the resonance of its first thickness mode ($f_0 = 1022 \,\mrm{kHz}$). Near its resonance, we found that its inverse impedance or admittance $Y=1/Z$ is described by a Lorentzian curve 
\begin{equation}
Y=\frac{1}{Z}=\frac{1}{R \left\{1+ \left[ Q \left( \omega - \omega_0 \right) /\omega_0 \right] ^2\right\} }\; ,
\end{equation}
where $\omega_0 = 2\pi f_0$. After measuring $R$, we fitted the above equation to obtain the quality factor $Q$ in helium~3 or 4 at low temperature. We obtained the preliminary values $Q = 175 \pm 25$ for helium~3 and $Q= 135 \pm 10$ for helium~4. Unfortunately, the thickness mode resonance is close to another resonance which must be a high order flexion mode, so that we do not take this determination of $Q$ as reliable.

The quality factor is of particular interest for the short sinusoidal pulses that we are using. Indeed, it characterizes the transient time that is necessary to build up the amplitude of the oscillation of the transducer. At a frequency $f_0$, this transient time is $\tau_0=Q/(2\pi f_0)$. For the sake of simplicity, we will take the period as the time unit; in other words, a time $t$ corresponds to $x$ cycles through the relation $x=f_0 t$. Let $V_0$ be the amplitude of the sinusoidal voltage and $n$ the total number of cycles in the electrical burst; hence the burst duration is $n$ in our reduced units. During the pulse ($0\leq x\leq n$), the intantaneous voltage is $V(x) = V_0 \sin(2 \pi x)$.

In a steady regime, i.e., in the long time limit, the ceramic surface would oscillate with an amplitude
\begin{equation}
\zeta _0= \frac{V_0}{2 (\pi f_0)^{3/2}} \sqrt{\frac{Q}{MR}}\; ,
\label{eq:zeta(V)}
\end{equation}
This equation is derived\cite{theseLambare98} by writing
\begin{equation}
Q = \omega_0 \frac{E_\mrm{stored}}{P_\mrm{diss}}\; ,
\label{eq:defQ}
\end{equation}
where $E_\mrm{stored}$ is the energy stored during one period and $P_\mrm{diss}={V_0}^2/(2 R)$ the average dissipated power. Let $l$ be the transducer thickness and $\zeta(z,t)$ the amplitude of the displacement at a distance $z$ from the inner surface. For the lowest thickness mode, which is a standing wave with one node at $z=l/2$, we have $\zeta(z,t)=\zeta_0 \cos(\pi z/l) \,\sin(\omega_0 t)$. The density of acoustic energy is the sum of a kinetic term and a pressure term that have the same amplitude. By integrating this energy density over the transducer volume and averaging over one period, we find the total acoustic energy $E_\mrm{acoust} = M {\omega_0}^2 {\zeta_0}^2/4$. For a piezoelectric transducer, there is an additional term in the stored energy:
\begin{equation}
E_\mrm{strored}=E_\mrm{acoust}+\frac{1}{4} \,C {V_0}^2 \frac{1}{1-{k_\mrm{t}}^2}\; ,
\label{eq:Estored}
\end{equation}
where $C$ is the transducer capacitance and $k_\mrm{t}$ the electromechanical coupling constant.
We measured $C=1 \,\mrm{nF}$ at low temperature. As for $k_\mrm{t}$, the constructor gives $0.47$ at room temperature, and an experimental check at low temperature gave $0.27$.\cite{theseLambare98} With these values and the measured resistance at resonance $R=12\,\mrm{\Omega}$, we find that the ratio between the second term in Eq.~(\ref{eq:Estored}) and $E_\mrm{stored}$ calculated from Eq.~(\ref{eq:defQ}) is of the order of $5\times 10^{-4}$. Therefore, we can neglect the second term of Eq.~(\ref{eq:Estored}) and write $E_\mrm{acoust}=E_\mrm{stored}$ to derive Eq.~(\ref{eq:zeta(V)}).

In the transient regime, $x$ cycles after the excitation has started, the actual surface oscillation is
\begin{figure}
\begin{center}
\psfig{file=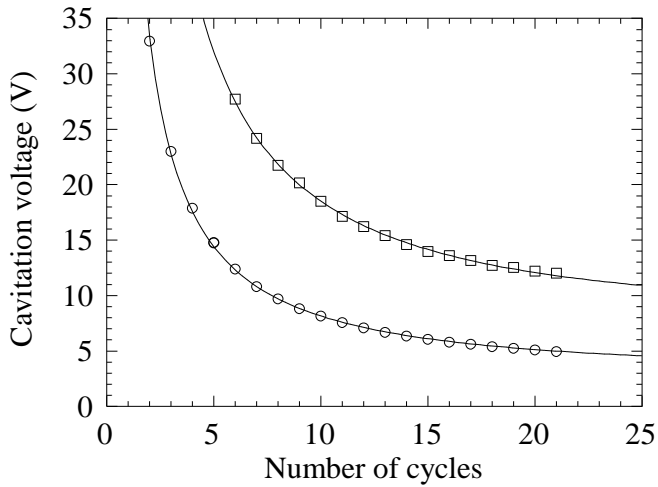,width=8.6cm}
\end{center}
\caption{Cavitation threshold voltage versus number of cycles in helium~3 (circles, $T=423 \,\mrm{mK}$ and $P=4 \,\mrm{mbar}$) and helium~4 (squares, $T=284 \,\mrm{mK}$ and $P=40 \,\mrm{mbar}$). The solid lines are fits with Eq.~(\ref{eq:Vc(n)}) which gives $Q=119$ in helium~3 and $Q=100$ in helium~4.}
\label{fig:Vc(n)}
\end{figure}
\begin{equation}
\zeta (x)=\zeta _0 \sin(2 \pi x)\left[1-\exp\left( -2 \pi \, \frac{x}{Q} \right) \right]
\label{eq:zeta(t)}
\end{equation}
for $0 \leq x \leq n$, and
\begin{equation}
\zeta (x)=\zeta _0 \sin(2 \pi x)\left[1-\exp\left( - 2 \pi \, \frac{n}{Q} \right) \right] \exp\left(-2 \pi \, \frac{x-n}{Q}\right)
\label{eq:zeta2(t)}
\end{equation}
for $x > n$. This leads to a better method for the measurement of the quality factor. We have used the dependence of the cavitation threshold voltage on the total number of cycles $n$ in the electrical pulse.

Indeed, cavitation occurs for a well defined value of the surface oscillation corresponding to a well defined pressure at the acoustic focus. It is the most negative pressure in the acoustic wave. Given the phase of the excitation voltage at time $t=0$, cavitation occurs for $x_\mrm{c}=n + \frac{1}{4}$, and we compared our measurements of the cavitation voltage $\Vc (n)$ to the equation
\begin{equation}
\Vc (n)= \frac{\exp\left(\frac{\pi}{2Q}\right)}{1-\exp\left(- 2 \pi \, \frac{n}{Q}\right)}\,\Vc^{\infty}\; .
\label{eq:Vc(n)}
\end{equation}

As shown on Fig.~\ref{fig:Vc(n)}, we obtained excellent fits with Eq.~(\ref{eq:Vc(n)}). In helium~3, we found $Q= 119 \pm 4$. In helium~4 we found $Q= 100 \pm 2$. In the same run, we also verified that the resonance frequency ($1022 \,\mrm{kHz}$) minimizes $\Vc$. We will use these values of $Q$ in the analysis below. We limited  the fit to $n \leq 21$ because for $Q=100$ and $n \geq 23$ cavitation occurs during the previous swing, at time $x_\mrm{c}=n - \frac{3}{4}$.

In principle, one expects slightly different values for $Q$ in helium~3 and in helium~4. The coupling of the ceramic to liquid helium is small, because the acoustic impedance $\rho c$ in liquid helium (of order $10^4\mrm{m^{-2}s^{-1}}$) is small compared to the impedance $\rho_\mrm{c} \, c_\mrm{c}$ of the ceramic ($3\times 10^{7} \,\mrm{m^{-2}\,s^{-1}}$). The ratio of the emitted power $P_\mrm{a}$ to the dissipated power $P_0$ in the ceramic is given by
\begin{equation}
\frac{P_\mrm{a}}{P_0} = \frac{4}{\pi}\frac{\rho \, c}{\rho_\mrm{c} \, c_\mrm{c}} \, Q\; .
\label{eq:}
\end{equation}
As a consequence, the emitted acoustic power should be $6\,\%$ for helium~3 and $15\,\%$ for helium~4, in qualitative agreement with our measurement. As for the electrical impedance at resonance, we found $12.2 \,\mrm{\Omega}$ in helium~3 and $15.6 \,\mrm{\Omega}$ in helium~4.

Having chosen the values of the frequency ($1022 \,\mrm{kHz}$) and the pulse width (three to six cycles), we studied the temperature and pressure dependence of the cavitation threshold.

\subsection{Thermal relaxation}

To ensure reproducibility of the measurements, we had to check the time needed by the system to reach a new steady state after each adjustment of the temperature regulation. In order to do this, we used the following procedure. Starting from a low temperature ($65 \,\mrm{mK}$, for instance), we set the driving voltage to a value such that the probability was around $20\,\%$; then we warmed up to a new temperature ($10 \,\mrm{mK}$ higher, for instance): the reading of our thermometers was almost instantaneous, whereas the probability increased slowly. This is because small temperature gradients between the acoustic focal region and the thermometer take some time to vanish completely (see Sec.~\ref{par:dissip}). Then we adjusted the driving voltage to keep the probability around $50\,\%$, where the system is most sensitive to temperature drifts. We waited until the probability measurement fluctuated around a constant value. After several checks around different temperatures, we found the relaxation time to be less than $90\,\mrm{min}$ at low temperature in helium~3; we thus decided to wait for $90 \,\mrm{min}$ after each temperature change in the lowest temperature range. The relaxation time was found to be less than $30 \,\mrm{min}$ for helium~3 above $200 \,\mrm{mK}$ and less than $5\,\mrm{min}$ for  helium~4 at all temperatures, so that it was sometimes possible to work faster. In addition, we always waited several hours after each liquid helium or nitrogen transfer, because longer drifts were observed, particularly in the optical detection of the bubbles. The latter problems are associated with the thermal contraction of the refrigerator.

\subsection{Dissipation at low temperatures}
\label{par:dissip}

Working in helium~3 at low temperatures raises problems of temperature inhomogeneities. Unlike superfluid
\begin{figure}
\begin{center}
\psfig{file=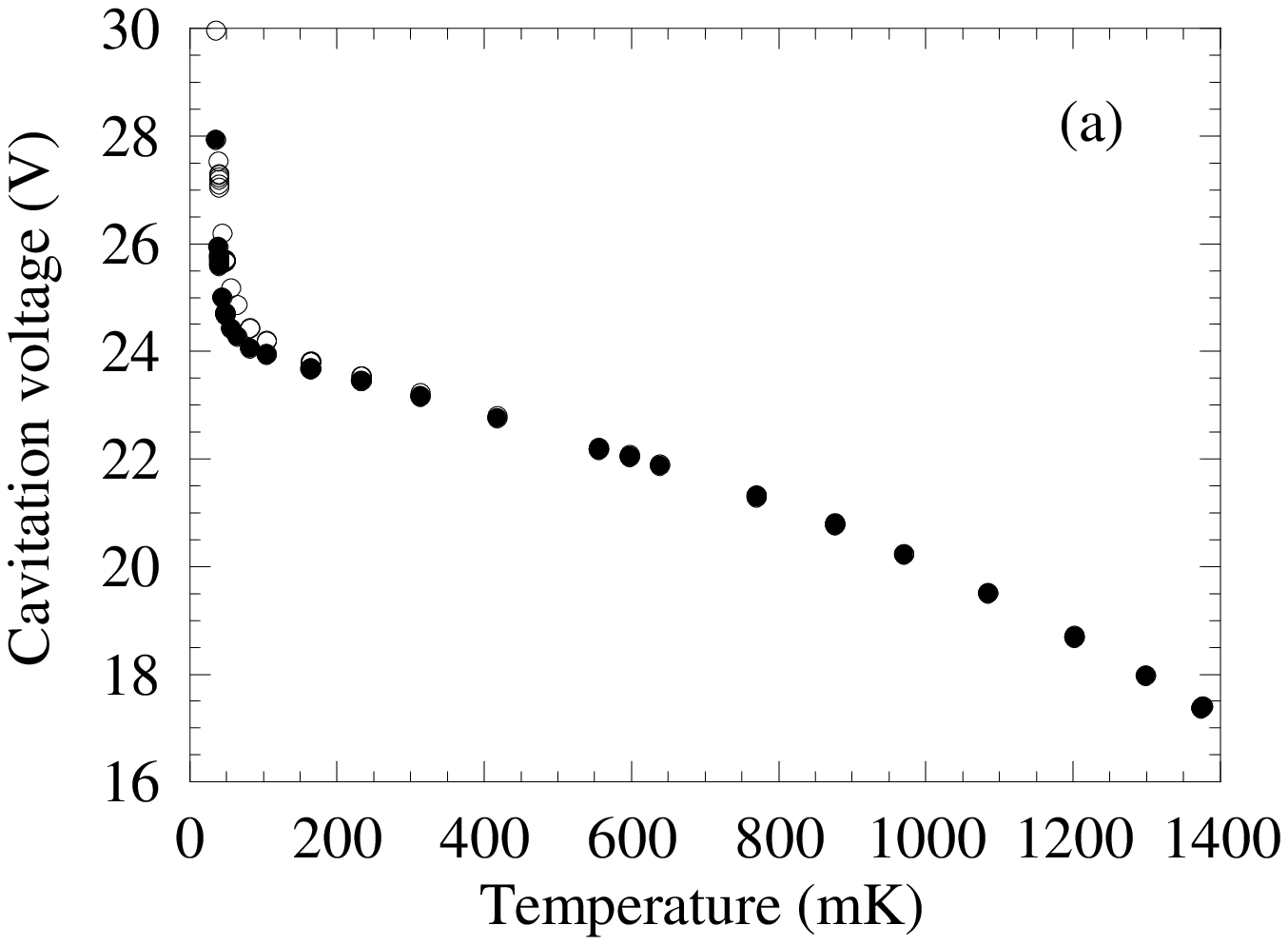,width=8.6cm}
\psfig{file=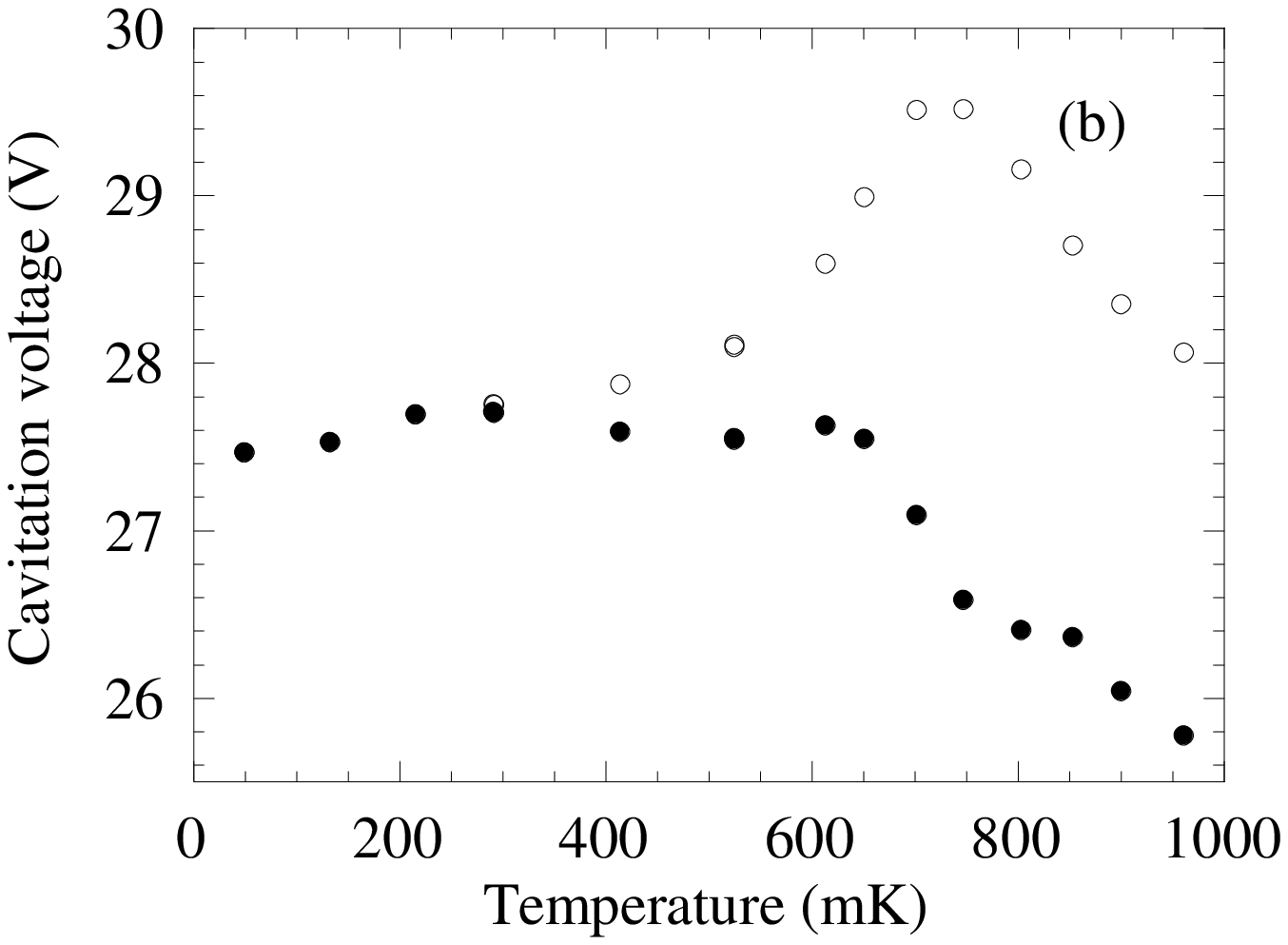,width=8.6cm}
\end{center}
\caption{Cavitation threshold voltage versus temperature: (a) in liquid helium~3 ($P=12 \,\mrm{mbar}$, $3 \,\mrm{cycles}$); (b) in liquid helium~4 ($P=37 \,\mrm{mbar}$, $6 \,\mrm{cycles}$). Empty circles represent raw data, and full circles data corrected from the sound attenuation.}
\label{fig:Vc(T)}
\end{figure}
{\noindent}helium~4, helium~3 is a poor thermal conductor. Therefore, the dissipation in the cell leads to temperature gradients between the cavitation area and the cell walls where the temperature is measured. There are two main sources of  dissipation: mechanical friction in the ceramic and absorption of the laser light in the cell. The former was minimized by decreasing the pulse duration and the repetition rate; the latter by attenuating the laser intensity and chopping the beam. We could estimate both effects by monitoring the heating power of the temperature regulation under different conditions. For the mechanical dissipation, we found $2.6 \,\mrm{\mu W}$ for six-cycle pulses repeated at $1 \,\mrm{Hz}$. We then chose to work with the shortest pulses allowed  by the power of the amplifier and the lowest affordable repetition rate, namely, three cycles and $0.05 \,\mrm{Hz}$. We checked that the cavitation voltage was independent of the repetition rate below $0.1 \,\mrm{Hz}$.

As for the light absorption we found it to be $0.7 \,\mrm{\mu W}$ in a cw operation mode, after having attenuated the laser beam. For a further reduction, we built an optical chopper that was synchronized with the acoustic pulse. With an ordinary electrical relay, we easily achieved an opening time of $20 \,\mrm{ms}$, so that the radiation power was divided by a factor of $1000$ for a repetition rate of $0.05 \,\mrm{Hz}$ and became completely negligible.

In helium~4 these problems did not exist and we worked with repetition rates ranging from $0.5$ to $2 \,\mrm{Hz}$ and a pulse width of six cycles. The lowest rates were used to reach the lowest temperatures; we kept the optical chopper for the same reason. We had to use longer pulses because cavitation requires a larger amplitude than in helium~3, and our rf amplifier has a limited output amplitude.

\subsection{Sound attenuation}

In all our experiments, we had to consider the attenuation of sound. In helium~3, this attenuation becomes important when the liquid enters the Fermi liquid region; this is below $100 \,\mrm{mK}$, where the viscosity varies as $1/T^2$ because of the temperature variation of the quasiparticle collision time.\cite{Wilks} We used the values of the viscosity measured by Bertinat~\etal\cite{Bertinat74} and the best fits they give in two temperature regions (Eqs.~(7) and (13) of Ref.~\ref{bib:Bert}).

In helium~4, the temperature variation of the sound attenuation shows a peak at the temperature where the phonon-roton collision time equals the sound period ($750 \,\mrm{mK}$ for a $1 \,\mrm{MHz}$ sound wave).\cite{Wilks} We used the absorption at $1 \,\mrm{MHz}$ extrapolated by Maris\cite{Marisprivate} from the measurements of Waters~\etal\cite{Waters67} at higher frequency.

In both cases, we assumed that, outside the focal region whose typical radius is one acoustic wavelength, the sound wave has a small enough amplitude to be treated as linear. The measured cavitation threshold voltage being related to the emitted amplitude, we assumed that the amplitude at the acoustic focus is reduced by the attenuation factor $\exp(-\alpha R_{\rm tran})$, where $R_{\rm tran}=8 \,\mrm{mm}$ is the transducer radius. We finally need to remark that in helium~4 we had checked this attenuation in our previous experiment, but in helium~3 a similar check appeared impossible to do. In helium~4, the value of the attenuation had been checked\cite{Lambare98} by measuring the amplitude of the light scattered by the acoustic wave. The temperature variation of this ``acoustic signal'' was found in good agreement with the known sound attenuation up to $0.9 \,\mrm{K}$; however, above $0.9 \,\mrm{K}$, the sound amplitude was found smaller than predicted by the sound attenuation only, as if additional mechanisms had to be considered also (diffraction, temperature drift of the transducer resonance frequency, etc.). In helium~3, a similar check would have required too much light amplitude and an average over too many bursts to be made in a reliable way in the interesting temperature region, which is in the low temperature limit.
\begin{figure}
\begin{center}
\psfig{file=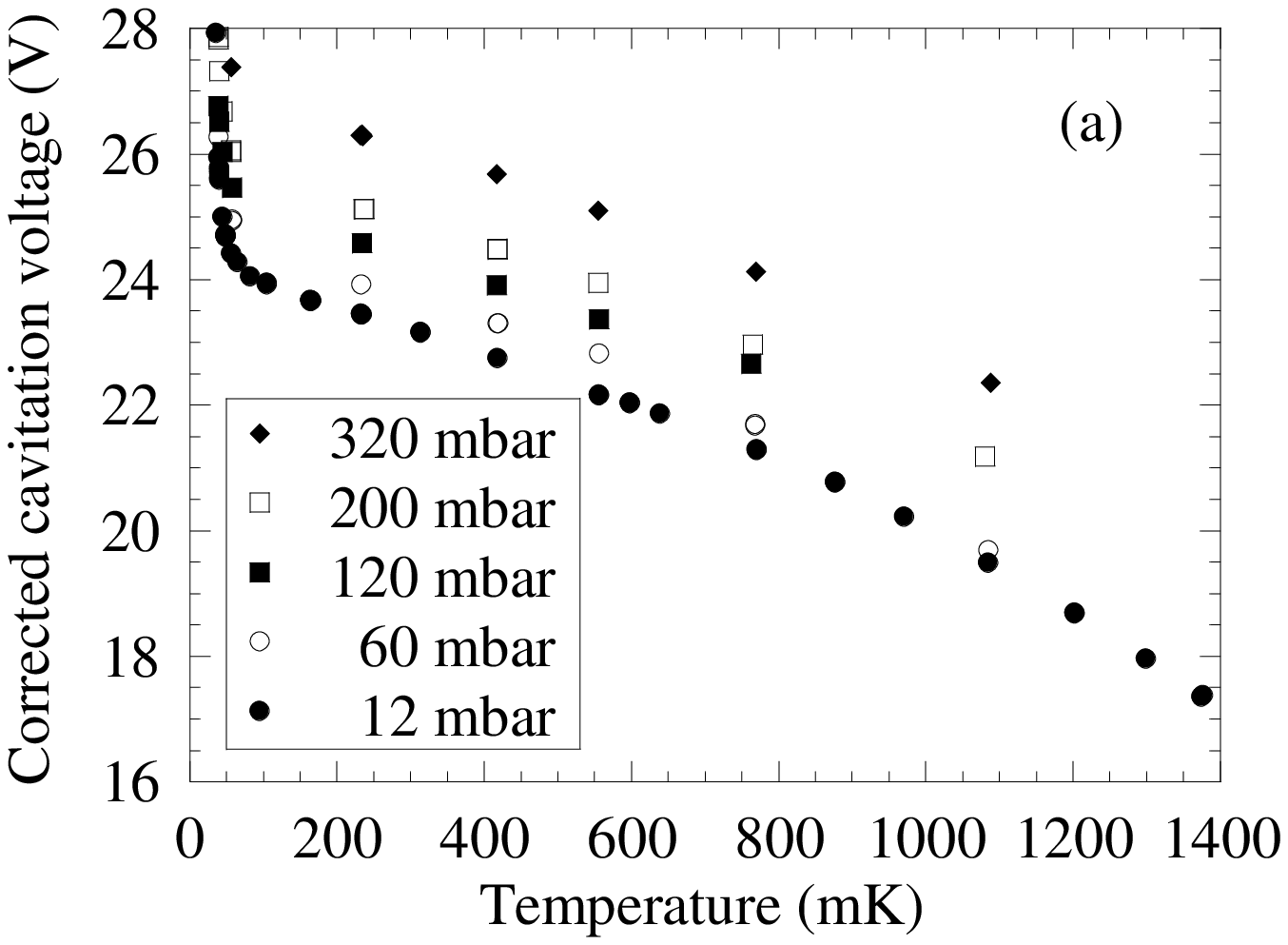,width=8.6cm}
\psfig{file=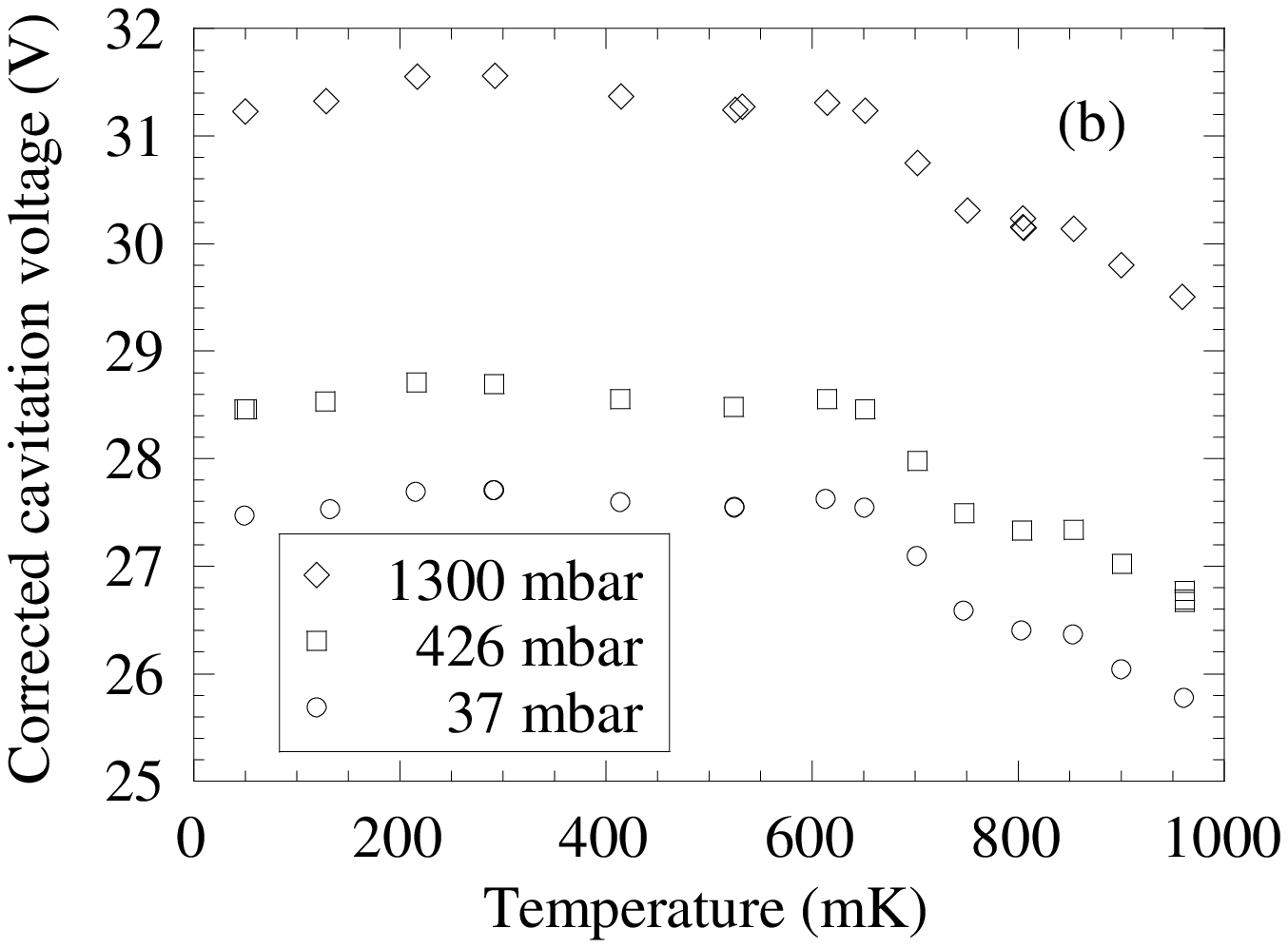,width=8.6cm}
\end{center}
\caption{Corrected cavitation threshold voltage versus temperature at different static pressures: (a) in liquid helium~3 ($3 \,\mrm{cycles}$); (b) in liquid helium~4 ($6 \,\mrm{cycles}$).}
\label{fig:Vc_P(T)}
\end{figure}

\subsection{Data corrections}

Let us start with helium~4. Figure~\ref{fig:Vc(T)}(b) presents a set of results for the cavitation threshold voltage as a function of the static temperature $T$ in the cell. The results obtained confirm what was already published by Lambar\'e~\etal\cite{Lambare98} There are two sets of data (raw data and corrected data). Fitting cavitation probability measurements with Eq.~(\ref{eq:S(V)}) gives the raw value of the threshold voltage $\Vc$; it is the excitation amplitude at which the probability is $0.5$. The correction accounts for the attenuation of the acoustic wave over its flight distance, i.e., the transducer radius $R_{\rm tran}$. The corrected data indicate the existence of a crossover around $T = 0.6  \,\mrm{K}$. Below $0.6 \,\mrm{K}$, we attribute the temperature independent regime to quantum cavitation occurring by quantum tunneling. Above $0.6 \,\mrm{K}$, the cavitation threshold decreases with increasing $T$, as expected for a thermally activated, classical regime. Our goal is to show that the cavitation threshold in the low temperature limit corresponds
\begin{figure}
\begin{center}
\psfig{file=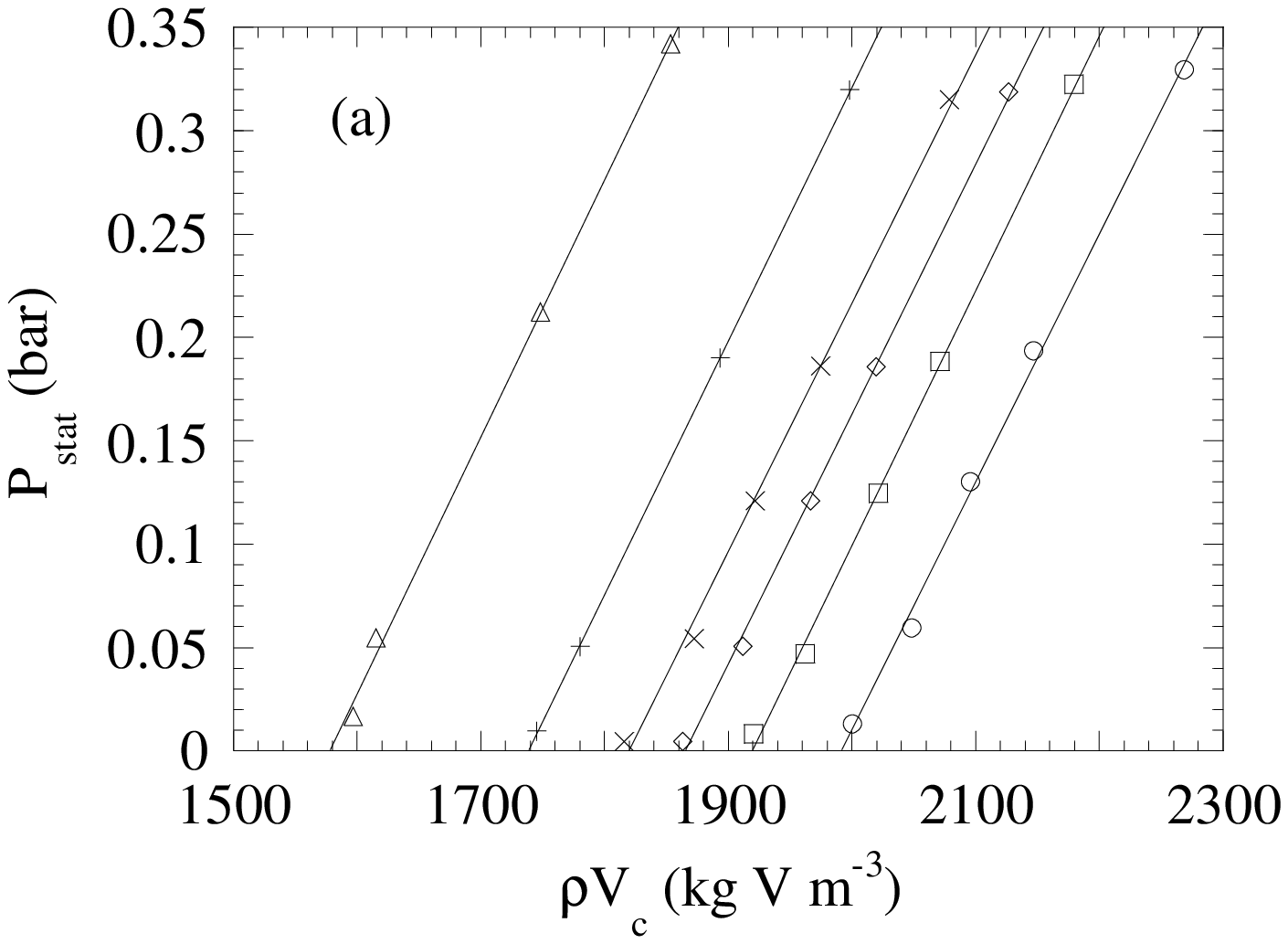,width=8.6cm}
\psfig{file=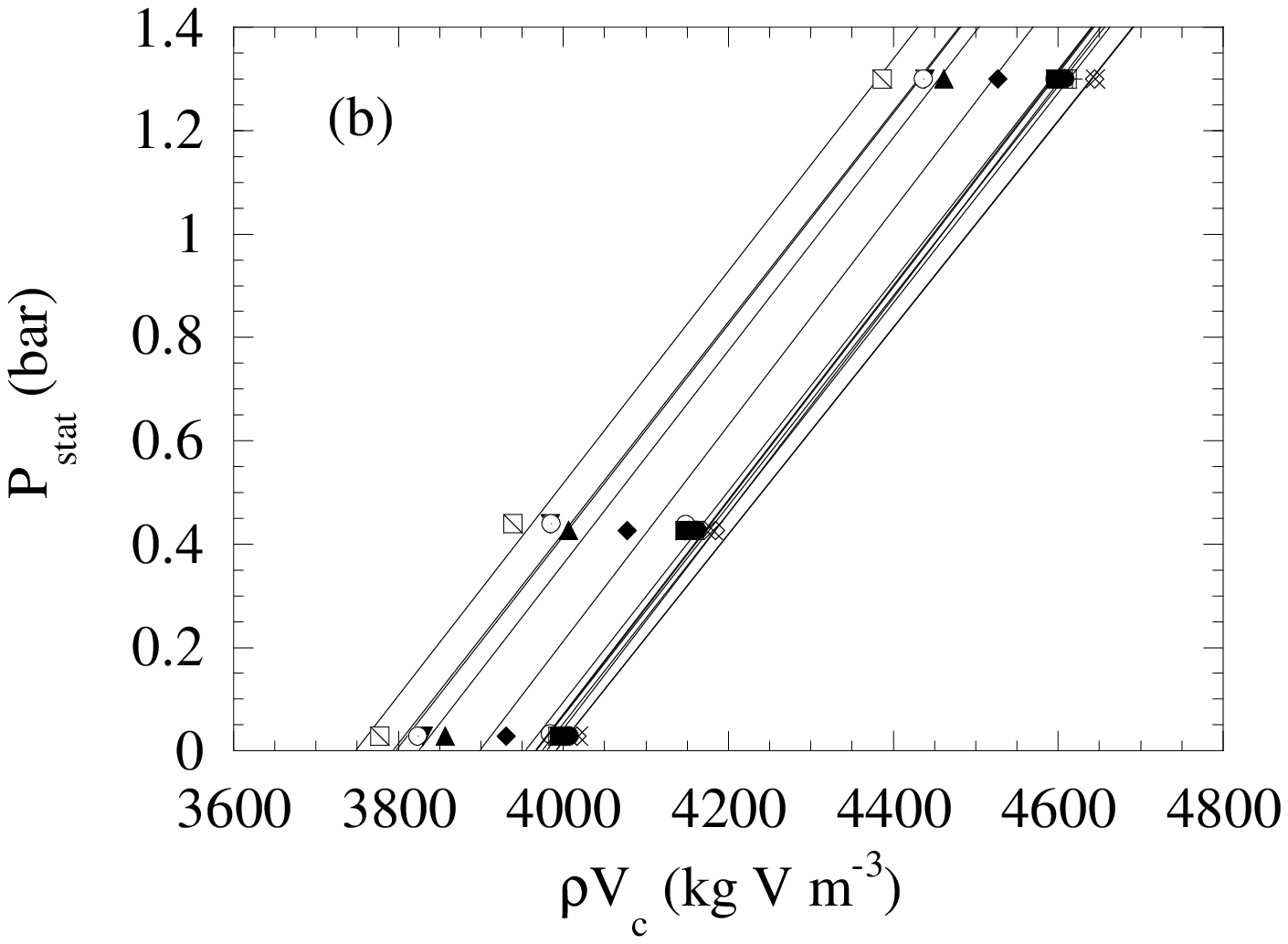,width=8.6cm}
\end{center}
\caption{Static pressure as a function of the parameter $\rho \Vc$ (see text): (a) in liquid helium~3 ($3 \,\mrm{cycles}$); (b) in liquid helium~4 ($6 \,\mrm{cycles}$). The pressures obtained by linear extrapolation are given in Table~\ref{tab}.}
\label{fig:Pstat(rhoVc)}
\end{figure}
{\noindent}to a pressure close to the spinodal limit $\Ps = -9.5 \,\mrm{bar}$. As explained in Ref.~\ref{bib:Lambare98}, agreement with the quantum cavitation theory of Maris requires that this quantum plateau corresponds to $-9.2 \,\mrm{bar}$. We have tried to check this value more precisely in this work.

As for helium~3, our results are presented in Fig.~\ref{fig:Vc(T)}(a). We have extended to lower temperatures the measurements we reported before.\cite{Caupin00} A striking difference from helium~4 is the sharp increase of the threshold in the very low temperature limit (below $55 \,\mrm{mK}$). It occurs in a temperature domain where the applied correction diverges. Indeed, the attenuation of sound is proportional to $T^{-2}$ in the Fermi liquid region. One possible origin of the experimental divergence could be that we do not apply the right correction. However, we do not believe so. If this effect is not an experimental artifact, it is possible to propose an interpretation for it by considering the particular properties of a Fermi liquid near its spinodal limit.\cite{Caupin01Fermi} In this article, we restrict ourselves to the temperature domain above $55 \,\mrm{mK}$ with the following question: does the cavitation pressure approach the calculated spinodal pressure $\Ps = - 3.15 \,\mrm{bar}$ there?

\section{Measurements under pressure}

We now turn to the dependence on the static pressure in the cell. If there was no attenuation of sound in liquid helium, if the coupling of the ceramic to the helium was independent of pressure, and if the sound amplitude at the focus was simply proportional to the driving voltage, then the estimation of the cavitation pressure would be very simple. By measuring the cavitation threshold voltage $\Vc$ as a function of the static pressure, we would observe a linear variation and then extrapolate down in pressure. The negative pressure at which $\Vc$ = 0 would be the pressure at which cavitation occurs in our experiment. Although things are not that simple, mainly because, in helium, the sound velocity strongly depends on pressure so that the focusing of sound is nonlinear, we have used the pressure dependence of the cavitation threshold as explained below.

\subsection{Results}

We repeated our measurements at different pressures using an identical set of temperatures for each pressure. We were limited to a few bars only because, as the pressure increases, bubbles become more and more difficult to detect. Indeed, as expected from the Rayleigh-Plesset theory\cite{Rayleigh17} and previously measured in helium~4,\cite{Roche96} the maximum radius of bubbles varies as ${\epsilon_0}^{1/3} {\Pst}^{-1/3}$, where $\epsilon_0$ is the energy acquired by the bubble from the acoustic wave. In helium~3, there is less acoustic energy available to make the bubble grow after its nucleation. Above $0.5 \,\mrm{bar}$ in helium~3 and 2 bar in helium~4, the bubbles are too small to be detected at the cavitation threshold. Our results are shown in Fig.~\ref{fig:Vc_P(T)}.

A correction has been applied to these data, which accounts for the sound attenuation. In order to do this, we first noticed that, in both helium~3  and helium~4, the viscosity does not significantly vary with pressure from $0$ to $3 \,\mrm{bar}$. For example, in helium~3, the viscosity $\eta$ varies as $T^{-2}$ and the low temperature limit of $\eta \, T^2$ varies by less than $6\,\%$ in this pressure range.\cite{Wheatley75} Furthermore, in this pressure study, there is only one measurement (namely, at $57 \,\mrm{mK}$) for which attenuation is not negligible; we discarded measurements below $55 \,\mrm{mK}$ in helium~3 because of their very large dependence on temperature which induces some scatter in the results and makes the pressure extrapolation imprecise.

In helium~4, the pressure dependence of the attenuation is not well known. Several groups have shown that the  reduced viscosity decreases and the peak temperature increases with increasing pressure. However, these effect are small. From the measurements of Dransfeld~\etal\cite{Dransfeld58} we estimated the relative variation of the viscosity as $-2.4\rm\,\%\,{bar}^{-1}$ and that of the attenuation peak temperature as $+1.8 \rm\,\%\,{bar}^{-1}$. Accordingly, we used a single value for the viscosity $\eta$ in our whole study.

Using Ref.~\ref{bib:Wilks} and references therein, we checked the temperature dependence of the density and of the sound velocity in our temperature ranges. The temperature variation of the density is smaller than $0.4 \,\%$ in helium~3 and smaller than $0.03\,\%$ in helium~4. The temperature variation of $c$ in helium~4 is less than $0.2\,\%$. As a consequence we neglected these temperature variations in our analysis (for more details, see Sec.~\ref{par:discuss}).

The only significant variations are the pressure variation of the density and that of the sound velocity. We took them into account in deriving the absorption coefficient:
\begin{equation}
\alpha=\frac{8 \pi^2 f^2 \eta}{3 \, \rho \, c^3}\; .
\label{eq:alpha}
\end{equation}
For the calculation of $\alpha$, we used the equation of state by Maris. For the present purpose, it would make no significant difference to use the one given in the Appendix.

We can now explain how we used the dependence of the cavitation voltage on the static pressure to measure the cavitation pressure.

\subsection{An upper bound for the cavitation pressure}
\label{par:upbound}
If the focusing of the sound wave was linear, the pressure swing at the focus would be
\begin{equation}
\Delta P = \rho \, \omega^2 R_{\rm tran} \zeta\; ,
\label{eq:linregime}
\end{equation}
where $\zeta$ is the displacement of the transducer wall. This displacement itself is proportional to the voltage applied to the ceramic since the ceramic oscillations are small enough to be in the linear regime. As a consequence, we have plotted the static pressure as a function of the product $\rho \Vc$; again, to calculate $\rho$ for the different pressures, we used the equation of state of Maris and we neglected the temperature variation. The result is shown in Fig.~\ref{fig:Pstat(rhoVc)}(a) for helium~3 and Fig.~\ref{fig:Pstat(rhoVc)}(b) for helium~4. We then extrapolated our measurements linearly to $\rho \Vc$ = 0. We claim that this extrapolation gives an upper bound for the cavitation pressure.

\begin{figure}
\begin{center}
\psfig{file=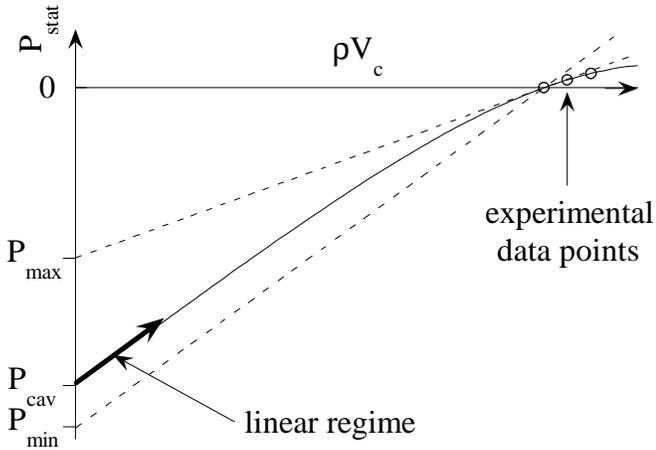,width=8.6cm}
\end{center}
\caption{Illustration of the method used to obtain bounds for the cavitation pressure (see text). For the sake of clarity, this figure is not drawn on scale.}
\label{fig:concav}
\end{figure}

Of course, if we knew the nonlinear relation between the pressure oscillation at the focus and the driving voltage $V$, we could directly obtain the value of the cavitation pressure; but we do not know this nonlinear relation: we only guessed it as shown by the solid line in Fig.~\ref{fig:concav}. One thing we know is the qualitative effect of nonlinearities. Suppose that one could work with a negative static pressure in the cell, close to the cavitation pressure. A small amplitude sound wave would then be sufficient to produce cavitation, and the focusing of this small amplitude wave would be in its linear regime, as described by
{\noindent}Eq.~(\ref{eq:linregime}). Since we cannot start from a negative static pressure, we are forced to use a large amplitude sound wave. To make a given pressure swing we need to drive the transducer with a larger voltage than if the regime were linear. This is because, as one reaches more and more negative pressures at the focus, the liquid there is more and more compressible, so that the excitation is less and less efficient in building up a negative pressure swing. If we considered the effects of nonlinearities on the magnitude of the positive pressure swing, it would be the opposite. 
We thus expect the large amplitude sound oscillation at the focus to be asymmetric, that is, to have a smaller negative swing amplitude and a larger positive swing amplitude than a sine wave. As we increase the static pressure, we need an even larger sound amplitude to reach the cavitation threshold, so that the nonlinear effects are larger and larger. As a consequence, we measure a slope $\partial \Pst/\partial(\rho \Vc)$ that is smaller than if we had no nonlinear effects; this corresponds to the negative curvature of the solid line in Fig.~\ref{fig:concav}. A linear extrapolation thus leads to a pressure less negative than the actual pressure at the focus; this means that an upper bound for $\Pcav$ is obtained. This effect of nonlinearities has been confirmed by the preliminary results of the recent numerical calculations by Appert~\etal\cite{Appert01} and by a very recent experimental study.\cite{Chavanne01} As shown in Fig.~\ref{fig:Pstat(rhoVc)} and in Table~\ref{tab}, we repeated this extrapolation for series of data at different temperatures; at low temperature, we found $-2.4 \,\mrm{bar}$ for helium~3 and $-8 \,\mrm{bar}$ for helium~4. These two bounds are only slightly larger than the calculated spinodal limits of these two liquids ($-3.15$ and $-9.65 \,\mrm{bar}$), in very good agreement with our expectations. 

\subsection{A lower bound for the cavitation pressure}

The same reasoning also leads to a lower bound for the cavitation pressure. Indeed, we can try to estimate the displacement $\zeta$ and use Eq.~(\ref{eq:linregime}) to obtain the magnitude of the pressure swing at the focus; the calculated value will be an overestimate of the actual magnitude. According to Eq.~(\ref{eq:zeta(V)}), when starting from a static pressure $\Pst$ and a temperature $T$, the pressure swing $\Delta P(\Pst, T)$ required to produce cavitation is proportional to $\rho(\Pst, T) \Vc^{\infty}(\Pst, T)$, where $\Vc^{\infty}(\Pst, T)$ is the cavitation voltage extrapolated to infinite pulse duration using Eq.~(\ref{eq:Vc(n)}). More precisely, we write
\begin{equation}
\frac{\Delta P(\Pst, T)}{\rho(\Pst, T) \Vc^{\infty}(\Pst, T)} = 2 \, R_{\rm tran} \sqrt{\frac{\pi f_0 Q}{MR}} \; .
\label{eq:slope}
\end{equation}
In the linear approximation, we have
\end{multicols}
\begin{center}
\begin{table}
\mediumtext
\caption{Upper and lower bounds for the cavitation pressure in helium~3 and helium~4 at several temperatures. These values are used in Fig.~\ref{fig:encadPcav}}.
\begin{tabular}{cccccc}
\multicolumn{3}{d}{Helium~3}&\multicolumn{3}{d}{Helium~4}\\
\tableline
Temperature&$\Pmax$&$\Pmin$&Temperature&$\Pmax$&$\Pmin$\\
(mK)&(bar)&(bar)&(mK)&(bar)&(bar)\\
\tableline
56.8&-2.39&-3.00&49.8&-8.06&-10.35\\
234&-2.36&-2.86&129&-8.06&-10.37\\
418&-2.25&-2.78&216&-7.98&-10.44\\
555&-2.19&-2.71&292&-7.98&-10.44\\
768&-2.13&-2.60&414&-8.08&-10.40\\
1085& -1.96&-2.37&525&-8.21&-10.38\\
&&&614&-8.28&-10.41\\
&&&652&-8.22&-10.38\\
&&&702&-8.14&-10.21\\
&&&749&-7.89&-10.02\\
&&&804&-7.77&-9.95\\
&&&854&-7.73&-9.93\\
&&&901&-7.69&-9.81\\
\end{tabular}
\label{tab}
\end{table}
\end{center}
\widetext
\begin{multicols}{2}
\begin{equation}
\Delta P(\Pst, T) = \Delta P(0, T) + \Pst\; ,
\end{equation}
so that $\Pst$ is a linear function of the product $\rho \Vc^\infty$ with a slope given by Eq.~(\ref{eq:slope}).

By inserting our measured values of $R$ and $Q$ in Eq.~(\ref{eq:slope}), we estimate the slope of the linear regime to be $1035 \;\mrm{Pa\;kg^{-1}\;V^{-1}\;m^3}$ in helium~3 and $899 \;\mrm{Pa\;kg^{-1}\;V^{-1}\;m^3}$ in helium~4. Now drawing a straight line with this calculated slope through the low static pressure data points, we find the lower bound for $\Pcav$ as the intersection with the vertical axis ($\rho \Vc^\infty=0$); Fig.~\ref{fig:concav} illustrates this construction. We repeated this procedure for series of data at different temperatures; at low temperature, we found $-3.0 \,\mrm{bar}$ in helium~3 and $-10.4 \,\mrm{bar}$ in helium~4.

The bounds obtained with this second method are more negative than the bounds obtained in Sec.~\ref{par:upbound}; this is equivalent to the fact that the experimental slopes of Sec.~\ref{par:upbound}, after extrapolation to infinite pulse duration, are smaller ($845 \pm 16 \;\mrm{Pa\;kg^{-1}\;V^{-1}\;m^3}$ in helium~3 and $660 \pm 15 \;\mrm{Pa\;kg^{-1}\;V^{-1}\;m^3}$ in helium~4) than the ones we calculated here. Given the experimental difficulties that we mentioned above, this is rather satisfactory and it supports our whole analysis.

\subsection{Discussion}
\label{par:discuss}

\begin{figure}
\begin{center}
\psfig{file=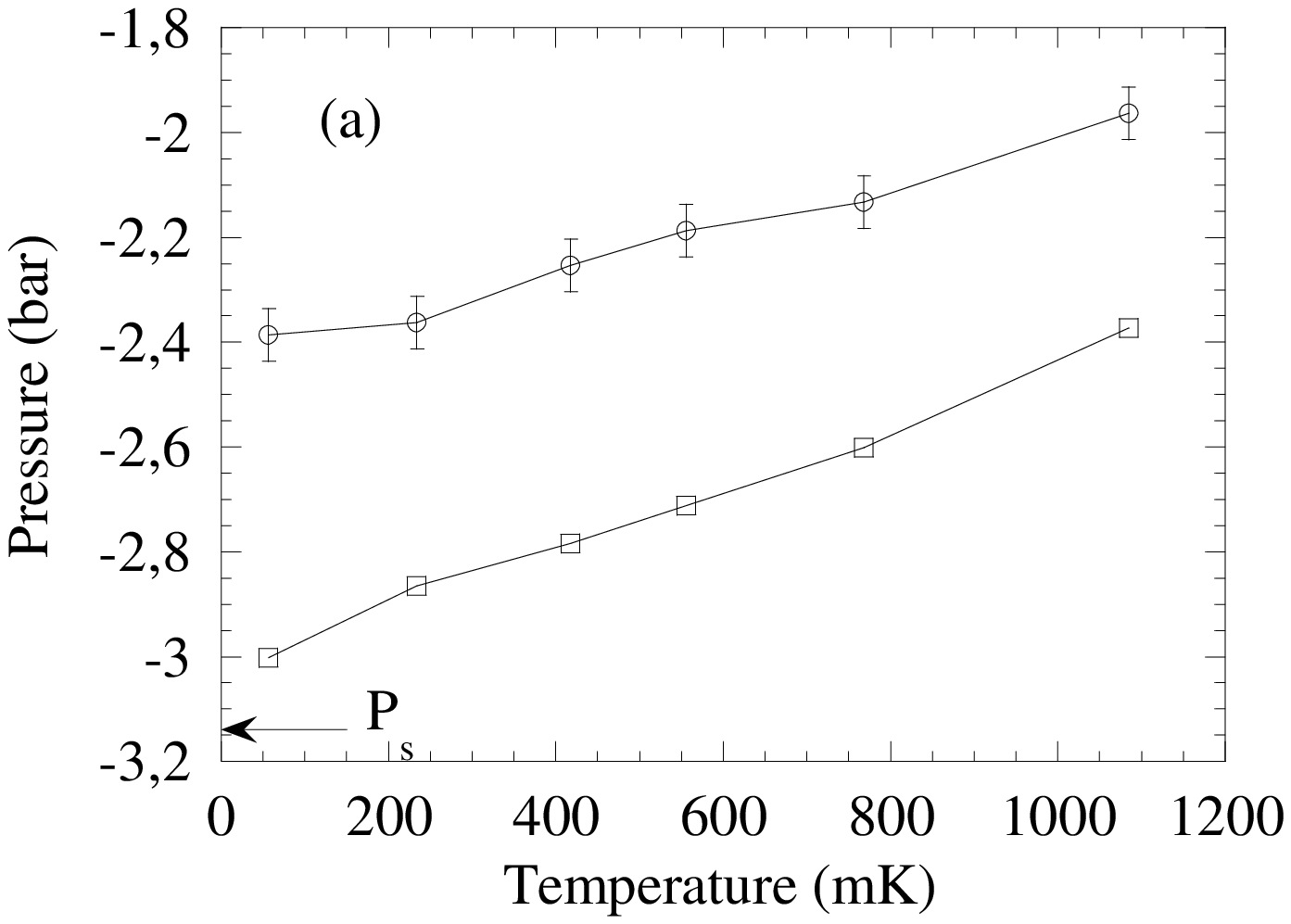,width=8.6cm}
\psfig{file=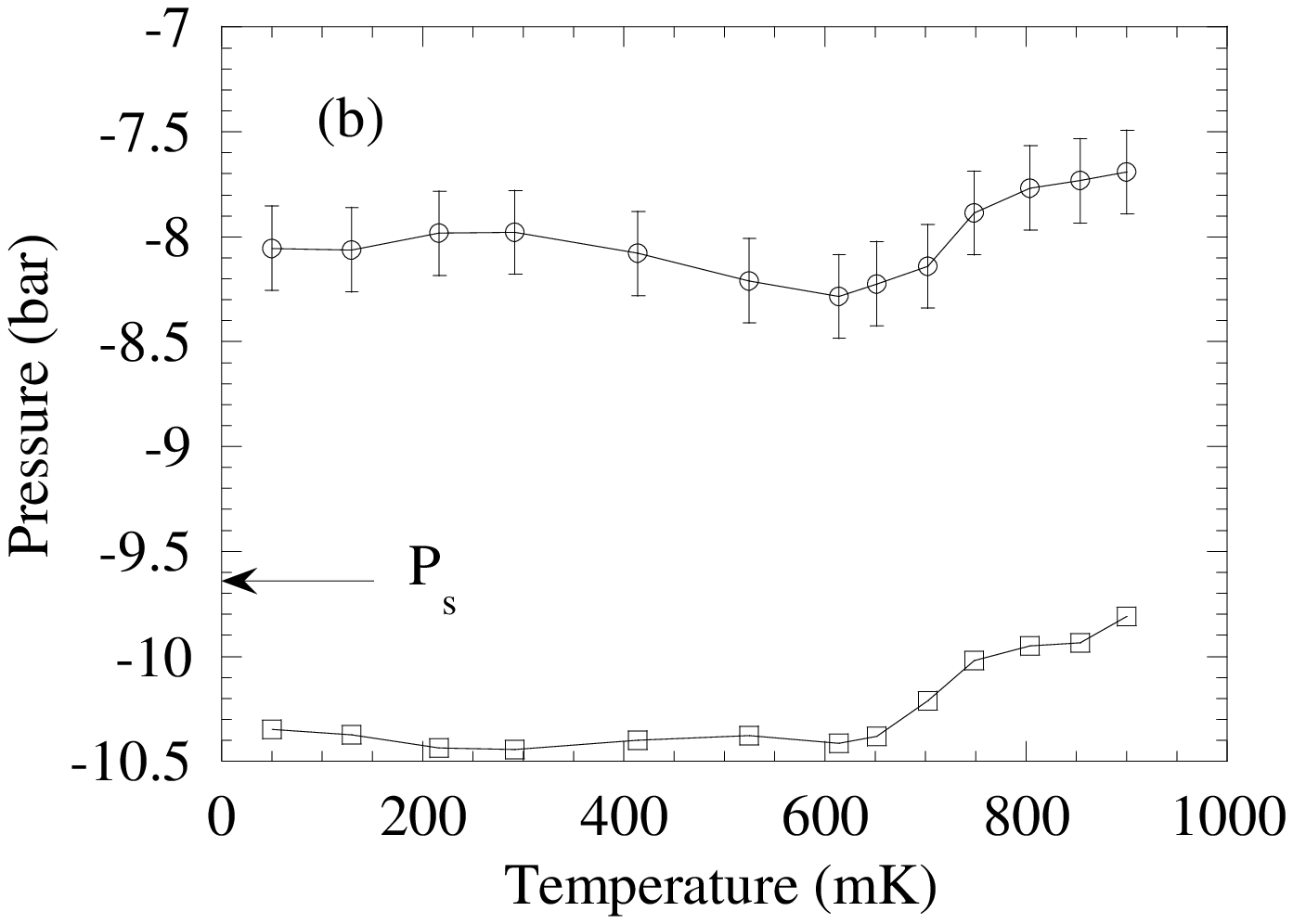,width=8.6cm}
\end{center}
\caption{Bounds obtained for the cavitation pressure in helium~3~(a) and helium~4~(b). Upper bounds are given by the circles and lower bounds by the squares. The arrows indicate the theoretical values of the cavitation pressure at low temperature. Error bars are discussed in Sec.~\ref{par:discuss}; their magnitudes for the lower bounds are smaller than the marker size.}
\label{fig:encadPcav}
\end{figure}

As now shown in Fig.~\ref{fig:encadPcav}, we have obtained experimental bounds for the cavitation pressure in helium~3~(a) and in helium~4~(b). We have estimated the error bars on the data points of Table~\ref{tab} and Fig.~\ref{fig:encadPcav}. Let us start with $\Pmax$. The uncertainty is about $\pm 0.05 \,\mrm{bar}$ for helium~3 and $\pm 0.2 \,\mrm{bar}$ for helium~4. For helium~3, we have tried to include the temperature dependence of the density by using Kollar and Vollhardt's analysis and program;\cite{Kollar} the effect is to lower all the points by less than $30 \,\mrm{mbar}$. As for $\Pmin$, the main uncertainty comes from the determination of the quantities $Q$ and $R$, which are used to calculate the slope from Eq.~(\ref{eq:linregime}). This may lead to a systematic error of about $\pm 5\,\%$, i.e., $\pm 0.15 \,\mrm{bar}$ in helium~3 and $\pm 0.5 \,\mrm{bar}$ in helium~4. In the determination of $\Pmin$, there is also a small uncertainty of $\pm 0.2\,\%$ from the measurement of $\rho \Vc$.

Within about $10\,\%$, our results agree with the calculated spinodal limits at low temperature. Of course, we confirm that helium~3 is about three times more fragile than helium~4, in the sense that cavitation occurs (the liquid breaks) at a pressure that is about three times less negative in helium~3 than in helium~4. In fact, what surprises us on this figure is the smallness of the difference between the upper bounds and the lower bounds. This difference is related to the magnitude of the nonlinear effects, and we find these nonlinear effects rather small. The smallness of the nonlinearities can also be seen in the fact that our measurements at different temperatures fall on parallel lines: the slope of $\Pst(\rho \Vc)$ is nearly constant although $\Vc$ varies. This is a somewhat surprising
{\noindent}behavior that would need to be confirmed by a direct measurement of the sound amplitude at the focus. However, with such an open geometry where there is nothing in the acoustic focal region, the optical measurement of the instantaneous sound amplitude seems very difficult to perform, as shown by the previous attempt by Nissen~\etal\cite{Nissen89} It would thus be very interesting to calculate these effects in our case of a hemispherical transducer. The only calculations performed up to now are done in a fully spherical geometry because it is one dimensional (everything depends on the distance $r$ to the center only). According to the existing calculations in this spherical geometry, the nonlinear effects are large.\cite{Appert01} A calculation in a hemispherical geometry looks much more difficult because it is two dimensional. Its results might be different because the local condition at the center is not the same: there is a velocity node at the center in the spherical geometry which is closed, and not in the hemispherical geometry which is open.

\section{Conclusion}

By studying the pressure dependence of cavitation in liquid helium, we have obtained bounds for the cavitation pressure: at low temperature, $-3.0 < \Pcav < -2.4 \,\mrm{bar}$ in helium~3 and $-10.4 < \Pcav < -8.0 \,\mrm{bar}$ in helium~4. These negative pressures are close to the calculated spinodal limits ($- 3.1 \,\mrm{bar}$ in helium~3 and $-9.5 \,\mrm{bar}$ in helium~4), as expected for homogeneous nucleation near absolute zero. In order to improve this accuracy, we believe that it is necessary to insert something in the cavitation region, for example a glass plate. However, in such a case, the plate may affect the cavitation conditions. This type of experiment is in progress in our laboratory.\cite{Chavanne01}
Our measurements also give the temperature dependence of the cavitation pressure: we will discuss it in a forthcoming paper.

\acknowledgments
We are grateful to H.~J.~Maris, H.~Lambar\'e, X.~Chavanne, C.~Appert, and D.~d'Humi\`eres for many useful discussions.

\appendix
\section*{}
The method used by Maris\cite{Maris94,Maris95} to obtain the value of the spinodal pressure $\Ps$ consists in extrapolating measurements of the sound velocity $c$ at positive pressure with a law of the form
\begin{equation}
c^3 = b \, (P - \Ps) \; .
\label{eq:c3(P)}
\end{equation}

Using the measurements of Abraham~\etal,\cite{Abraham70,Abraham72} Maris found the value of $\Ps$ to be $- 3.097 \,\mrm{bar}$ in helium~3 (Ref.~\ref{bib:Maris95}) and $-9.5219 \,\mrm{bar}$ in helium~4 (Ref.~\ref{bib:Maris94}). However, he took pressures in bars whereas they are in atmospheres in the original papers by Abraham~{\etal} To check the effect of
\begin{figure}
\begin{center}
\psfig{file=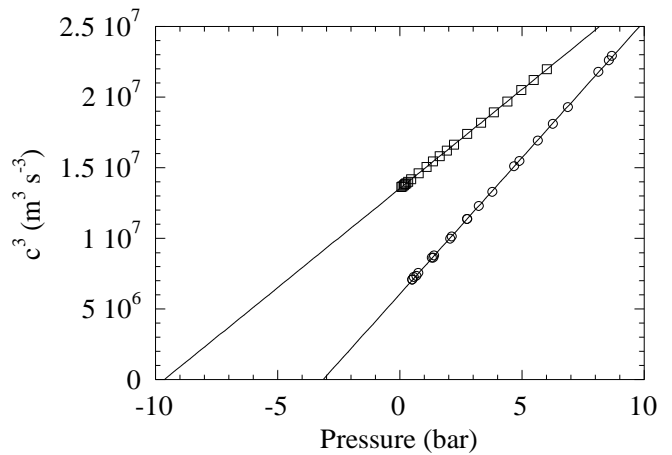,width=8.6cm}
\end{center}
\caption{Cube of the sound velocity as a function of pressure at $150 \,\mrm{mK}$ in helium~3 (circles) and helium~4 (squares). Linear fits (solid lines) give the spinodal pressure: $\Ps\simeq -3.14 \,\mrm{bar}$ in helium~3 and $\Ps\simeq -9.65 \,\mrm{bar}$ in helium~4.}
\label{fig:c3(P)}
\end{figure}
{\noindent}this mistake on $\Ps$, we performed the extrapolation with Eq.~(\ref{eq:c3(P)}) after the appropriate unit conversion. We used the same set of data points as did Maris, namely, the ones ranging from $0$ to $10 \,\mrm{atm}$ in helium~3 and from $0$ to $6 \,\mrm{atm}$ in helium~4. The linear fit of $c^3$ shown in Fig.~\ref{fig:c3(P)} gives $-3.1371 \,\mrm{bar}$ in helium~3 and $-9.6456 \,\mrm{bar}$ in helium~4 for the spinodal pressure $\Ps$.

Since the experimental error bar is on $c$ rather than on $c^3$, we find it better to fit the data of Abraham~\etal with the following formula:
\begin{equation}
c = \left[b (P - \Ps)\right]^{1/3} \; .
\label{eq:c(P)}
\end{equation}
We now obtain $\Ps=- 3.1534 \,\mrm{bar}$ in helium~3 and $-9.6435 \,\mrm{bar}$ in helium~4, and $b=19.262 \;\mrm{m^4\;s^{-1}\;kg^{-1}}$ in helium~3 and $14.030 \;\mrm{m^4\;s^{-1}\;kg^{-1}}$ in helium~4. Note that we present the values with five digits in order to show the correction, although we think that the extrapolation does not attain this level of accuracy.
\end{multicols}


\begin{references}
\bibitem{Mariserror} We have found a slight mistake in the papers by Maris. He used bars instead of atmospheres, and we have corrected his values (see the Appendix).
\end{references}
\end{document}